\documentclass[fleqn,usenatbib]{mnras}

\usepackage{newtxtext,newtxmath}

\usepackage[T1]{fontenc}

\DeclareRobustCommand{\VAN}[3]{#2}
\let\VANthebibliography\thebibliography
\def\thebibliography{\DeclareRobustCommand{\VAN}[3]{##3}\VANthebibliography}

\usepackage{graphicx}	
\usepackage{amsmath}	
\usepackage{amssymb}	
\usepackage{float}
\usepackage{pdflscape}

\title[Y358 and Literature UDG Dynamical Masses]{Keck Spectroscopy of the Coma Cluster Ultra-Diffuse Galaxy Y358: Dynamical Mass in a Wider Context}

\author[J. S. Gannon et al.]{Jonah S. Gannon$^{1}$\thanks{E-mail: jgannon@swin.edu.au},
Duncan A. Forbes$^{1}$,
Jean P. Brodie$^{1,2}$,
Aaron J. Romanowsky$^{3,2}$,
\newauthor{Warrick J. Couch$^{1}$
 and Anna Ferr\'e-Mateu$^{4,5}$}
\\
$^{1}$ Centre for Astrophysics and Supercomputing, Swinburne University, John Street, Hawthorn VIC 3122, Australia
\\
$^{2}$ University of California Observatories, 1156 High Street, Santa Cruz, CA 95064, USA
\\
$^{3}$ Department of Physics and Astronomy, San Jos\'e State University, One Washington Square, San Jose, CA 95192, USA
\\
$^{4}$ Instituto de Astrof\'isica de Canarias, Calle V\'ia L\'actea S/N, E-38205, La Laguna, Tenerife, Spain
\\ 
$^{5}$ Departamento de Astrofísica, Universidad de La Laguna, 38206, La Laguna (S.C. Tenerife), Spain
\\
}

\date{Accepted XXX. Received YYY; in original form ZZZ}

\pubyear{2022}

\begin{document}
\label{firstpage}
\pagerange{\pageref{firstpage}--\pageref{lastpage}}
\maketitle

\begin{abstract}
We examine ultra-diffuse galaxies (UDGs) and their relation to non-UDGs in mass--radius--luminosity space. We begin by publishing Keck/KCWI spectroscopy for the Coma cluster UDG Y358, for which we measure both a recessional velocity and velocity dispersion. Our recessional velocity confirms association with the Coma cluster and Y358's status as a UDG. From our velocity dispersion (19 $\pm$ 3 km s$^{-1}$) we calculate a dynamical mass within the half-light radius which provides evidence for a core in Y358's dark matter halo. We compare this dynamical mass, along with those for globular cluster (GC)-rich/-poor UDGs in the literature, to mass profiles for isolated, gas-rich UDGs and UDGs in the NIHAO/FIRE simulations. We find GC-poor UDGs have dynamical masses similar to isolated, gas-rich UDGs, suggesting an evolutionary pathway may exist between the two. Conversely, GC-rich UDGs have dynamical masses too massive to be easily explained as the evolution of the isolated, gas-rich UDGs. The simulated UDGs match the dynamical masses of the GC-rich UDGs. However, once compared in stellar mass -- halo mass space, the FIRE/NIHAO simulated UDGs do not match the halo masses of either the isolated, gas-rich UDGs or the GC-rich UDGs at the same stellar mass. Finally, we supplement our data for Y358 with other UDGs that have measured velocity dispersions in the literature. We compare this sample to a wide range of non-UDGs in mass--radius--luminosity space, finding UDGs have a similar locus to non-UDGs of similar luminosity with the primary difference being their larger half-light radii. 
\end{abstract}




\section{Introduction} \label{sec:intro}

The class of ``ultra-diffuse galaxy'' (UDG) was first coined by \citet{vandokkum2015} in relation to a subset of large half-light radius, low surface brightness galaxies in the Coma cluster. Formally, they classified UDGs as galaxies with half-light radius, $R_{\rm e} > 1.5$ kpc, and central surface brightness, $\mu_{0,g}>24\ \mathrm{mag\ arcsec^{-2}}$. Galaxies fitting this definition have been discovered in a wide range of environments both before (e.g., \citealp{Disney1976, Sandage1984, Bothun1987, Impey1988, Impey1997, Dalcanton1997}) and after 2015 (e.g., \citealp{Yagi2016, MartinezDelgado2016, vanderBurg2017, Roman2017, Roman2017b, Roman2019, Janssens2017, Janssens2019, Muller2018, Prole2019, Forbes2019, Forbes2020b, Zaritsky2019, Zaritsky2021, Barbosa2020}).

It is worth noting that since their coining by \citet{vandokkum2015}, many other authors have applied the same UDG term to galaxies fitting a different criteria set. For example, in the catalogue of \citet{Yagi2016} the UDG size criterion was relaxed to $R_{\rm e}>0.7$ kpc. Other authors have used a surface brightness criterion based on the average surface brightness within the half-light radius ($\langle\mu\rangle_{\rm e}$; e.g., \citealp{vanderBurg2017, Janssens2017, Janssens2019, Gannon2021b}), or altered the surface brightness/filter band at which it applies (e.g., \citealp{Janssens2017, Janssens2019, Forbes2020b}). The studies of \citet{RuizLara2019} and \citet{Chilingarian2019} went further, applying the UDG term to a set of galaxies that are generally brighter and smaller than the original definition. Using the ROMULUS simulations, \citet{vanNest2022} found that the choice of UDG criteria is key, having a large impact on the implied mechanisms underpinning their formation. Specifically, definitions for what comprises a `UDG' that are less restrictive may dilute the link between objects fitting the definition and their underlying formation mechanism.


For UDGs, their necessarily faint nature means spectroscopy of their stellar body requires a large time investment on 8m+ class telescopes. While gas-rich UDGs can be studied using their gas-kinematics (e.g., \citealp{ManceraPina2019, ManceraPina2022, Kong2022}) this method is not available for quiescent UDGs. As such, spectroscopy has largely focused on deriving the properties of single, or a small handful of, UDGs (e.g., \citealp{vanDokkum2017, Toloba2018, Alabi2018, Ferremateu2018, MartinNavarro2019, Emsellem2019, Danieli2019, vanDokkum2019b, Muller2020, Gannon2020, Gannon2021, Gannon2021b, Forbes2021}). Many of these UDGs have been targeted because of their extreme properties, even within the UDG class. For example many UDGs have received targeted spectroscopy due to their anomalously populous globular cluster (GC) systems (e.g., Dragonfly 44 and DFX1 \citealp{vanDokkum2017, vanDokkum2019b}), a known indicator of a massive dark matter halo \citep{Spitler2009, Harris2017, forbes2018, Burkert2020, Zaritsky2022}. While this has led to a slew of interesting discoveries, it has likely resulted in an overall literature that is poorly representative of the UDG population as a whole.

Simulations of galaxy formation primarily propose that UDGs form in a `puffy dwarf' scenario. In brief, they suggest UDGs are simply an extension of the regular dwarf galaxy population to larger sizes. The primary cause of this puffing up is usually attributed to higher than average halo spin \citep{Amorisco2016, Rong2017, Liao2019}, strong stellar feedback \citep{DiCintio2017, Chan2018}, tidal forces/quenching \citep{Carleton2019, Sales2020, Tremmel2020}, early mergers \citep{Wright2021} or combinations of the aforementioned four \citep{Jiang2019, Martin2019, Liao2019}. It seems likely that these scenarios account for many, perhaps even most, galaxies residing in the UDG definition.

Early work incorporating GCs into simulations of `puffy dwarf' UDG formation suggested the formation of GC-rich UDGs may be possible in dwarf-like dark matter halos \citep{carleton2021}. However, this formation scenario cannot explain known GC--dark matter halo mass scaling relations and is unable to produce GC-rich UDGs in massive dark matter halos \citep{Gannon2021b}.

Alternatively, it has been suggested that GC-rich, massive halo UDGs may be the dark matter dominated remnants of the earliest phases of galaxy formation. The observational expectation is for the galaxy to have quenched early and catastrophically. In doing so it fails to form a large portion of its expected stellar mass (\citealp{vanDokkum2016, Peng2016, Villaume2022, Danieli2022, Janssens2022, Buzzo2022}). These massive halo UDGs are not reproduced by leading cosmological simulations of galaxy formation. We note the work of \citet{Saifollahi2021, Saifollahi2022} which suggested the rich GC systems of 5 previously studied Coma cluster UDGs may be the result of measurement error. However even after their measurement corrections, \citet{Saifollahi2022} concluded an early formation and quenching scenario is still one of the most viable formation pathways. \citet{Saifollahi2022} referred to this UDG formation process as a `failed dwarf galaxy' scenario. Furthermore, follow-up spectroscopy, which allows measurement of a dynamical mass, largely supports the idea that GC-rich UDGs may reside in massive dark matter halos \citep{vanDokkum2019b, Gannon2020, Forbes2021, Gannon2021b}. It is not currently clear what fraction of the population massive halo UDGs represent.

For pressure-supported systems it has been well established that in velocity dispersion, effective radius and surface brightness space, galaxies reside on a so-called ``fundamental plane" \citep{Djorgovski1987, Faber1987, Dressler1987}. The fundamental plane offers unique insights into the physical processes generating pressure-supported systems and thus constrains their formation (e.g., \citealp{Borriello2003, Cappellari2006, Forbes2008, Graves2010, Tollerud2011, Zaritsky2019}). The fundamental plane and altered forms of it, such as mass--radius--luminosty space \citep{Tollerud2011} or the fundamental manifold \citep{Zaritsky2006}, have been shown to extend over nearly eight orders of magnitude in luminosity. These offer a connection from the dwarf spheroidals to giant elliptical galaxies \citep{Zaritsky2006, Forbes2008, Tollerud2011}. This allows an exploration of the relationship between luminous matter and dark matter halos from the smallest to the largest structures in the Universe. It is also critical to our understanding of the dominant galaxy formation processes on different mass scales. With detailed studies of mass profiles being prohibitively time intensive and still leaving great uncertainty in total UDG halo masses \citep{vanDokkum2019b}, placing large samples of UDGs on these relations is key to understanding their formation (cf., \citealp{Gannon2021b}). We adopt the latter approach in this work.  

Here we present new Keck II/Keck Cosmic Web Imager (KCWI) spectroscopy for the Coma cluster UDG Y358. From these data we measure both a recessional velocity and a velocity dispersion (Section \ref{sec:kcwi_data}). From our velocity dispersion we measure a dynamical mass. We compare this dynamical mass to dark matter mass profiles to look for evidence of a core or cusp (Section \ref{sec:results}). We additionally compare Y358's dynamical mass, along with dynamical masses for other UDGs, to mass profiles of isolated, gas-rich UDGs along with those from the NIHAO and FIRE simulations (Section \ref{sec:kong}). To contextualise this comparison we compare the UDGs, both observed and simulated, in stellar mass -- halo mass space (Section \ref{sec:smhm}). We then supplement our Y358 data with those for literature UDGs with the intention of placing all on the fundamental plane (Section \ref{sec:lit_sample}). In Section \ref{sec:biases} we discuss the biases present in our sample. In Section \ref{sec:MRL} we place UDGs in mass--radius--luminosity space, discussing their location on the plane in the context of UDG formation compared to non-UDGs on the plane. We present the concluding remarks of our study in Section \ref{sec:conclusions}. The literature sample discussed in Sections \ref{sec:MRL} \& \ref{sec:MRL} is presented in Appendix \ref{app:literature_sample}.







\begin{figure}
    \centering
    \includegraphics[width = 0.98 \columnwidth]{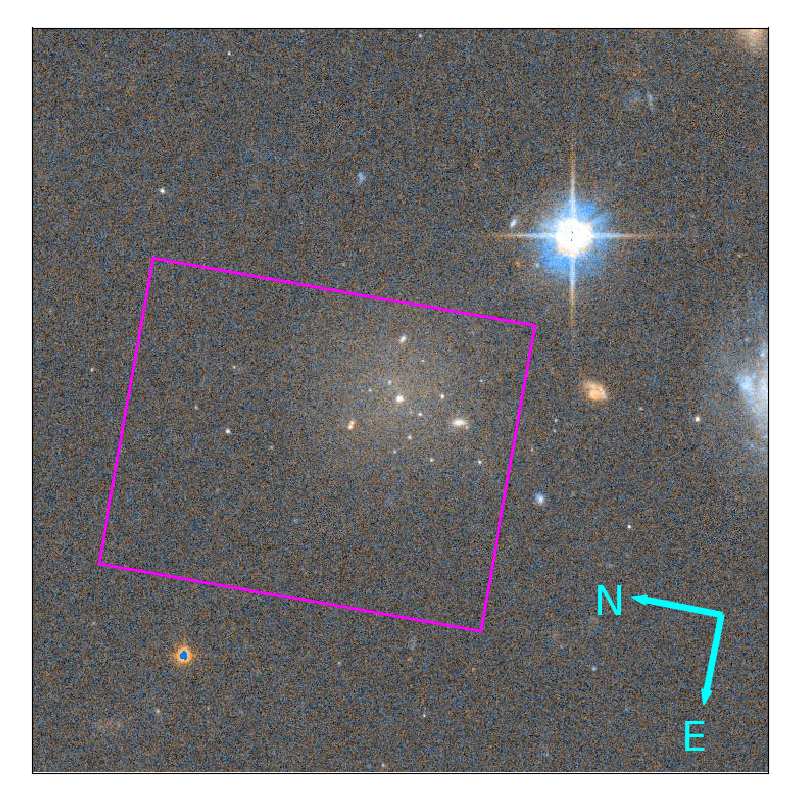}
    \caption{A 0.6' $\times$ 0.6' (17.5 kpc $\times$ 17.5 kpc at Coma Cluster distance), two colour (F814W/F475W) \textit{HST} cutout centred on Y358. The magenta rectangle indicates the positioning of the KCWI field of view. North and East are as indicated (cyan arrows). Of note are the numerous compact sources that appear associated with Y358, suggesting it likely hosts a rich GC system \citep{Lim2018}. A central compact source suggests it is nucleated.}
    \label{fig:hst}
\end{figure}

\begin{figure*}
    \centering
    \includegraphics[width = 0.98 \textwidth]{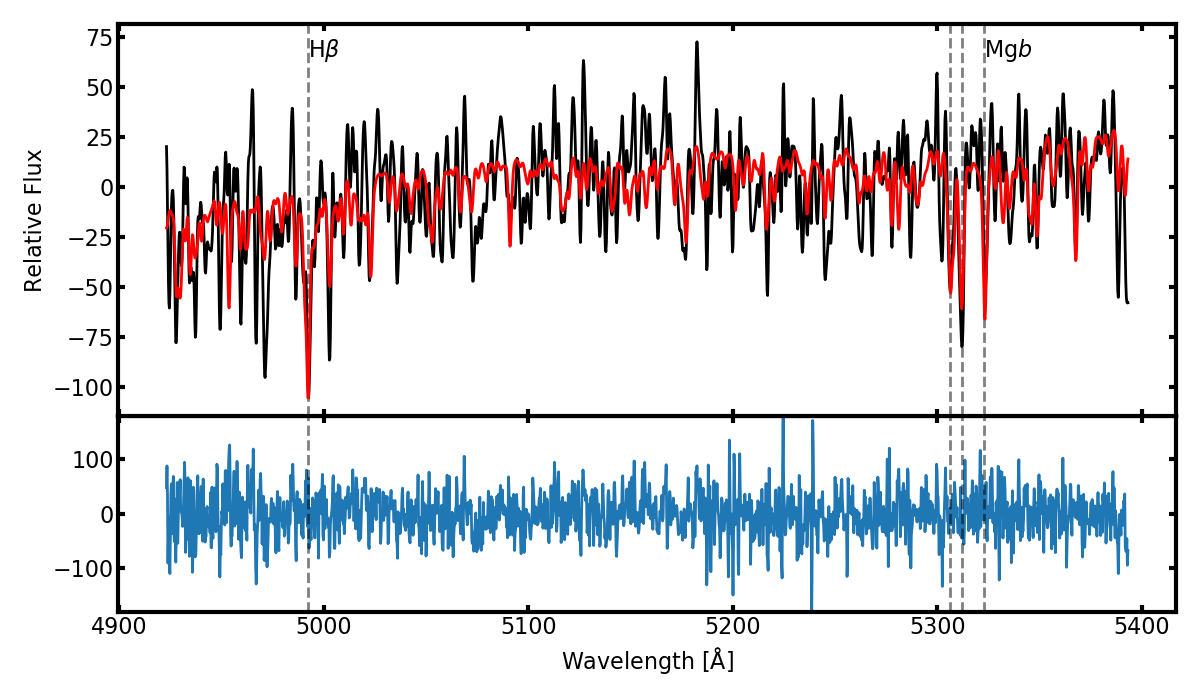}
    \caption{A Gaussian smoothed ($\sigma$ = 0.5 \AA) KCWI spectrum for Y358 (black) with example \texttt{pPXF} fit (red). Residuals from the non-smoothed fit are show at the bottom (blue). The spectrum, fit and residuals are displayed at the observed wavelengths. The prominent H$\beta$ and Mg$b$ triplet absorption features are indicated by dashed vertical lines.}
    \label{fig:spectrum}
\end{figure*}

\begin{figure}
    \centering
    \includegraphics[width = 0.98 \columnwidth]{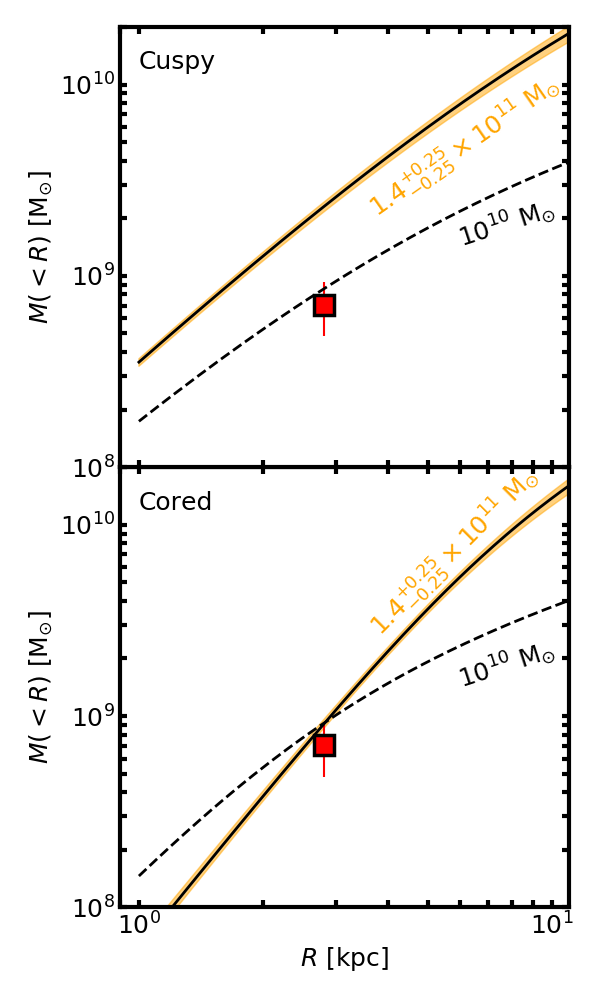}
    \caption{Enclosed mass \textit{vs.} galactocentric radius. We plot our dynamical mass measurement for Y358 (red square). In both panels we plot a halo profile of total mass expected from Y358's GC richness and the relationship of \citet{Burkert2020} (solid line with orange shading corresponding to the GC count uncertainty). We also include a halo profile of total mass roughly expected for a GC-poor UDG (i.e., $N_{\rm GC}$ = 2, $M_{\rm Halo}=$10$^{10}$ $\mathrm{M_{\odot}}$; dashed line). In the upper panel we plot these halo masses as cuspy NFW halos and in the lower panel we plot them as cored \citet{dicintio2014} halos. For Y358 to reside in a dark matter halo of mass expected from its GC counts, it likely resides in a cored dark matter halo.}
    \label{fig:y358_profile}
\end{figure}

\begin{figure}
    \centering
    \includegraphics[width = 0.98 \columnwidth]{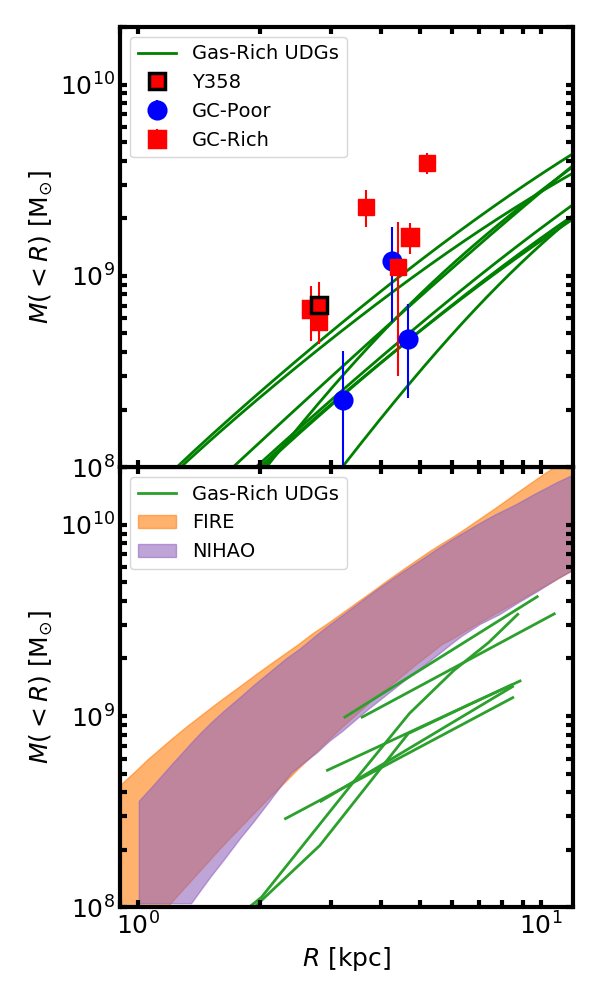}
    \caption{ Enclosed mass \textit{vs.} galactocentric radius. \textit{Upper:} Dynamical mass measurements for UDGs are plotted from \citet{Gannon2021b} (unbordered red and blue symbols) along with our Y358 dynamical mass measurement (black border). UDGs with rich ($N_{\rm GC}$~>~20) GC systems are plotted as red squares. UDGs with poor ($N_{\rm GC}$~<~20) GC systems are plotted as blue circles. The dark matter component of the best-fitting \citet{Read2016} halo profile to observations of gas-rich field UDGs are shown as green lines \citep[see their equation 4 and table 1]{Kong2022}. For the \citet{Kong2022} UDGs we exclude the gaseous component of their mass as our UDGs are dark matter dominated without gas. Dynamical mass measurements for GC-rich UDGs are on average too high for their formation to be easily explained as the transformation of the isolated, gas-rich \citet{Kong2022} UDGs. \textit{Lower:} The observed HI mass profiles from \citet{Kong2022} now include the mass from stars and gas (green lines). We include the range of halo mass profiles reproduced by the NIHAO simulations of \citet[purple shaded region]{DiCintio2017} and the FIRE simulations of \citet[orange shaded region]{Chan2018}. For NIHAO these profiles include the gas, stars and dark matter. The FIRE simulated mass profiles do not include gas as they artificially quench their galaxies as part of their simulation. Both the NIHAO simulations and the FIRE simulations predict mass profiles more massive than the observed UDGs of \citet{Kong2022}. This is despite the observed UDGs and the simulated UDGs being in isolated environments.}
    \label{fig:kong_compare}
\end{figure}

\section{New Keck Cosmic Web Imager Data} \label{sec:kcwi_data}
Here we present new KCWI data for the UDG, Y358. We target this galaxy due to its rich GC system which is indicative of a massive dark matter halo. Using the GC counts for Y358 from \citet{Lim2018} ($N_{\rm GC}$ = 28.0 $\pm$ 5.3) and the $N_{\rm GC}$ -- halo mass relationship of \citet{Burkert2020}, we infer a total dark matter halo mass of (1.4 $\pm$ 0.25) $\times 10^{11}$ $\mathrm{M_{\odot}}$ for Y358. The \citet{Lim2018} GC number is between the richness found by \citet[45$\pm$14]{vanDokkum2017} and the 90\%~ upper limit from \citet[18.4]{Amorisco2018} for Y358. This is also the number used in the study of \citet{Forbes2020} for Y358. Based on this GC richness, Y358 is expected to have a dark matter halo $>1$~$\sigma$ more massive than expected given its stellar mass ( $M_{\star} = 1.38\times 10^{8}$ $\mathrm{M_{\odot}}$; \citealp{Forbes2020}).

The integral field spectroscopy for the UDG Y358 was observed using KCWI (\mbox{\citealp{Morrissey2018}}) on 2020, March 21st (Program: U191; PI: Brodie). Skies were dark and clear with 1.2'' seeing. KCWI was configured using the medium slicer and `BH3' grating with a central wavelength of 5170 \AA\ (R $\approx$ 9900; $\sigma_{\rm inst}\ \approx$ 13 $\mathrm{km\ s^{-1}}$). We display a \textit{Hubble Space Telescope} (\textit{HST}) image of the galaxy, along with the KCWI pointing in Figure \ref{fig:hst}.

The data were reduced using the standard KCWI data reduction pipeline along with the extra post-pipeline trimming and flat fielding steps described in \citet{Gannon2020}. Spectra were extracted from the reduced data cubes using a 7 by 13 spaxel box centred on the galaxy with offset regions of the slicer as subtracted sky. These spectra were then barycentric corrected \citep{TollerudBary} and median combined. The resulting spectrum has S/N of 11 \AA$^{-1}$ with a total exposure time of 26400s. This spectrum has a wavelength range of 4923 \AA~to 5393 \AA.

We fitted the spectrum using \texttt{pPXF} \citep{Cappellari2017} and the \citet{Coelho2014} library with 241 different combinations of input parameters as per previous work (i.e., \citealp{Gannon2020, Gannon2021, Gannon2021b}). We display a smoothed version of our final spectrum, along with an example fit and fit residuals, in Figure \ref{fig:spectrum}. Our final values for the recessional velocity (7969 $\pm$ 2 km s$^{-1}$) and the velocity dispersion (19 $\pm$ 3 km s$^{-1}$) were taken from the median of these fits. We consistency checked these by fitting the red and blue halves of the spectrum. We also fitted the entire spectrum using a KCWI observation of the Milky Way globular cluster Messier 3 as a template. These consistency checks were all within the uncertainties for our quoted values for both recessional velocity and velocity dispersion. 

For the imaging properties of Y358 we use the values reported in Table 1 of \citet{vanDokkum2017}. We summarise the properties of Y358 in Table \ref{tab:sample}. Our recessional velocity for Y358 confirms its association with the Coma cluster. Combining this confirmation with the \citet{vanDokkum2017} imaging, we are able to confirm its status as a UDG. 


\section{Results} \label{sec:results}
We measure a dynamical mass for Y358 within the 3D de-projected half-light radius ($R_{1/2}$) using the mass estimator of \citet{Wolf2010}. Using the 2D projected, circularised half-light radius ($R_{e, \mathrm{circ}}$) and the luminosity-weighted line-of-sight velocity dispersion within this radius ($\sigma$), it takes the form: 

\begin{equation} \label{eqtn:wolf}
M(<R_{1/2}) = 930 \left( \frac{\sigma_{\rm e}^{2}}{\mathrm{\left(km\ s^{-1}\right)^{2}}}\right) \left(\frac{R_{\rm e, \mathrm{circ}}}{\mathrm{pc}}\right)\ \mathrm{M_{\odot};}\quad
\end{equation}
\begin{equation*}
\mathrm{where}\ R_{1/2} \approx \frac{4}{3} R_{\rm e, \mathrm{circ}}
\end{equation*}

We note this equation requires the luminosity-weighted line-of-sight velocity dispersion within the half-light radius. Our extracted region on Y358 corresponds to a $\sim 6.3'' \times 6.7''$ region which is only slightly smaller than the effective diameter ($\sim$ 9''). Our measured velocity dispersion of 19 $\pm$ 3 km s$^{-1}$ should well approximate the required value for Equation \ref{eqtn:wolf}. We therefore calculate a dynamical mass of 7.1 $\pm$ 2.2 $\times10^{8}$ $\mathrm{M_{\odot}}$ within 2.8 kpc for Y358 using Equation~\ref{eqtn:wolf}.


\begin{table*}
    \begin{tabular}{lllll}
    \hline
    Sample & Log($M_{\star}/$ $\mathrm{M_\odot}$) & Log($M_{\rm Halo}/$ $\mathrm{M_{\odot}}$) & Environment & Gas Content \\ \hline
    Observed: Y358 & $8.14$ & $11.06-11.22$ & Cluster & None \\
    Observed: GC-Rich & $8.04-8.89$ & $>11$ & Cluster/Group & None \\
    Observed: GC-Poor & $8.41 - 8.76$ & $<11$ & Cluster & None \\
    Observed: Kong+ (2022) & $7.45-8.35$ & $9.86-10.76$ & Field & Rich \\
    NIHAO: Di Cintio+ (2017) & $6.83-8.4$ & $10.22-10.85$ & Field & Rich \\
    NIHAO: Jiang+ (2018) & $6.8-8.8$ & $9.9-11.1$ & Field/Group & Rich \\
    NIHAO: Cardona-Barrero+ (2020) & $6.5-9.0$ & $10.04-11.29$ & Field & Rich \\
    FIRE: Chan+ (2018) & $7.72-8.44$ & $10.34-10.74$ & Field & None \\ \hline
    \end{tabular}%
    \caption{Pertinent properties of UDG samples relating to the discussion of Figure \ref{fig:kong_compare}. From left to right columns are: 1) Sample description. When relevant the simulation name is given before the literature reference; 2) Stellar mass range; 3) Halo mass range; 4) Environment; and 5) Gas content of the sample. The halo mass range of \citet{Cardona2020} was not published in that work and was provided upon request by the corresponding author.}
    \label{tab:comparison}
\end{table*}

\subsection{Y358 Halo Mass} \label{sec:dyn_mass}
In Figure \ref{fig:y358_profile} we compare the halo mass estimated from GC counts to the dynamical mass measurement we have obtained using our KCWI data. The comparison of a total halo mass to a mass measurement made within a fixed radius requires the assumption of a dark matter halo profile. Here, we assume a cuspy, NFW \citep{Navarro1996} halo profile along with a cored, \citet{dicintio2014} halo profile. We additionally plot a halo of mass roughly expected for a GC-poor UDG ($N_{\rm GC} = 2$; $M_{\rm Halo}$ =  10$^{10}$ $\mathrm{M_{\odot}}$; \citealp{Burkert2020}) for each of the cuspy/cored profiles. 

When the profile is forced to be a cuspy NFW profile with normal concentration (i.e., a concentration from \citealp{Dutton2014}) a halo of total mass $\sim 6.6 \times 10^{9}$ $\mathrm{M_{\odot}}$ is required to have the same enclosed mass as our measurement for Y358. Y358 residing in such a low mass dark matter halo is highly unexpected as it is below the \citet{Burkert2020} prediction from its rich GC system (i.e., 1.4 $\pm$ 0.25 $\times 10^{11}$ $\mathrm{M_{\odot}}$). There is evidence in the literature that UDGs should obey this relationship (e.g., \citealp{Gannon2021b}) therefore we suggest that Y358 does not reside in a low mass NFW halo. We instead conclude that Y358 likely resides in a cored and/or low concentration halo profile as it must do in order to obey the \citet{Burkert2020} relation. Previous works studying UDG dynamical masses have come to similar conclusions for other UDGs (e.g., \citealp{vanDokkum2019b, Gannon2021b}).



\subsection{Comparison to Gas-Rich UDGs} \label{sec:kong}

Recent work has suggested some isolated, gas-rich field UDGs may reside in low concentration and/or cored dark matter halos (see e.g., \citealp{Brook2021, ManceraPina2022, Kong2022}). Using resolved HI kinematics for seven such gas-rich isolated UDGs, \citet{Kong2022} were able to fit \citet{Read2016} mass profiles. Their best-fitting parameters are listed in their table 1. \citet{Read2016} mass profiles have the benefit of being able to reproduce observed dark matter cores in the dwarf halo mass regime while providing a convenient fitting function for star/gas kinematics.

We plot these best fitting \citet{Read2016} mass profiles from \citet{Kong2022} for comparison to UDG stellar kinematics in Figure \ref{fig:kong_compare} \textit{upper}. We note that our GC-rich UDGs are primarily in the cluster environment and hence gas poor. The contribution of baryons to their calculated dynamical mass is therefore small ($\lessapprox10$\%) and they are extremely dark matter dominated in their dynamics. In contrast for the gas rich UDGs, the gaseous component contributes significantly to the centrally enclosed mass (see e.g., the total masses in \citealp{Kong2022} table 1). We therefore choose to compare our data to the \citet{Read2016} mass profiles for the gas-rich UDGs as these trace the dark matter component of the halo which are more appropriate to compare to the measurements we are getting for our GC-rich/poor UDGs.


In Figure \ref{fig:kong_compare} \textit{upper} we plot our dynamical mass measurement for Y358 along with dynamical mass measurements for UDGs with stellar velocity dispersions and GC counts from \citet{Gannon2021b}. The UDGs in this sample are generally expected to be older and gas-poor due to their association with clusters. Only one of these UDGs is not associated with a cluster (i.e., NGC~5846\_UDG1) and it is in a group environment. UDGs from the \citet{Gannon2021b} sample have stellar masses in the range 8.04 < log($M_{\star}$ / $\mathrm{M_{\odot}}$) < 8.89. Four of the 7 \citet{Kong2022} UDGs have stellar masses in this range, with the remaining 3 having stellar masses slightly smaller (i.e., 7.45 < log($M_{\star}$ / $\mathrm{M_{\odot}}$) < 8.35). Plotted mass measurements are colour coded by GC richness based on a rich/poor divide of $N_{\rm GC}$~$\ge$~20 / $N_{\rm GC}$~$<$~20. The halo mass implied for a GC-rich UDG with $N_{\rm GC}$~$\ge$~20 is $\ge$10$^{11}$ $\mathrm{M_{\odot}}$ \citep{Burkert2020}. We summarise the pertinent properties (i.e., stellar mass, halo mass, environment and gas content) of these two observational samples in Table \ref{tab:comparison}.

It is clear from Figure \ref{fig:kong_compare} \textit{upper} that the GC-rich UDGs have dynamical masses that are too high to agree with the best fitting dark matter halos from \citet{Kong2022}. We note that this may be a reflection of the different total halo masses of the two UDG populations. i.e., the best-fitting total halo masses of the isolated, gas-rich UDG in \citet{Kong2022} are all below 10$^{10.8}$ $\mathrm{M_{\odot}}$ which is less than the minimum inferred halo mass for a GC-rich UDG with $N_{\rm GC}$>20 (10$^{11}$ $\mathrm{M_{\odot}}$; \citealp{Burkert2020}). Additionally, gas-rich UDGs tend to be younger, bluer with more irregular morphologies than other UDGs \citep{Leisman2017}, likely indicating ongoing star formation. Furthermore, recent work has shown that isolated, gas-rich UDGs do not have rich GC systems \citep{Jones2022}. We conclude that gas-rich UDGs similar to those observed by \citet{Kong2022} could not evolve into the GC-rich UDGs observed at present times. The progenitors of GC-rich UDGs require more massive dark matter halos at fixed stellar mass. 


This conclusion is not true for the GC-poor UDGs plotted in Figure \ref{fig:kong_compare}. All three of these UDGs have dynamical masses in agreement with the mass profiles of \citet{Kong2022}. We suggest it is possible that GC-poor UDGs in clusters have similar dark matter halo characteristics to isolated, gas-rich UDGs. Further, this suggests the processing and passive evolution of isolated, gas-rich UDGs is a possible formation pathway for GC-poor UDGs in clusters. This conclusion is similar to proposals from previous works (see e.g., \citealp{Roman2017, Martin2019, Grishin2021}). Our results therefore support GC-rich UDGs forming in more massive dark matter halos than GC-poor UDGs, with GC-poor UDGs being the possible evolution of isolated, gas-rich UDGs.

In Figure \ref{fig:kong_compare} \textit{lower} we plot the mass range of UDG profiles modelled in the NIHAO simulations of \citet[purple band]{DiCintio2017} and in the FIRE simulations of \citet[orange band]{Chan2018} \textit{vs.} the observed isolated, gas-rich UDGs from \citet[green lines]{Kong2022}. Both simulations primarily model UDGs as `puffy dwarfs' with large sizes driven by strong supernovae feedback. Both of the simulations are restricted to modelling UDGs in a relatively isolated environment, similar to the environment of \citet{Kong2022}'s observations. Additionally, the stellar mass range and total halo mass range of the UDGs modelled in the simulations provides good coverage of the stellar mass range and best-fitting total halo mass range of \citet{Kong2022}'s observed UDGs. We note, however, the recent observational work of \citet{Kadofong2022} which found that isolated, gas-rich UDGs do not exhibit the bursty star formation histories expected from these simulations. We summarise the stellar mass, total halo mass, environment and gas-richness of each sample in Table \ref{tab:comparison}. With similar environments, stellar masses and total halo masses, we might expect these simulations to reproduce the mass profiles of the observed isolated, gas-rich UDGs.

To make this comparison, in Figure \ref{fig:kong_compare} \textit{lower} we now plot the observed HI mass profiles from \citet{Kong2022}, which include both the gaseous and stellar component of the mass along with the dark matter. The \citet{DiCintio2017} NIHAO result plotted is also a total mass profile, including stars, gas and dark matter. Note that the FIRE mass profiles do not include gas, as they artificially quench their UDGs as part of their simulation. Without this artificial quenching their UDGs may still be expected to be gas-rich at present times. For both the \citet{DiCintio2017} and \citet{Kong2022} data plotted in Figure \ref{fig:kong_compare} \textit{lower}, the gas mass is, on average, more massive than the stellar component.

It is clear from Figure \ref{fig:kong_compare} \textit{lower} that both simulations create mass profiles more massive than the isolated, gas-rich UDGs that their simulations are best matched to reproduce. This is despite having a similar total halo mass. The mass profiles from the simulations are instead more closely matched to the GC-rich UDGs plotted in the \textit{upper} panel. However, few GC-rich UDGs have been observed in the low density environments such as those simulated, with some authors suggesting environment plays a key role in their GC formation \citep{Prole2019b, Somalwar2020}. Furthermore, based on their GC-richness and the GC number -- halo mass relationship of \citet{Burkert2020}, the GC-rich UDGs plotted are all expected to reside in halos of total mass greater than either the FIRE or NIHAO simulations at the same stellar mass. In order to have the observed GC-rich UDGs residing in halos with the total mass that is modelled in the simulations at the same stellar mass, they cannot follow the \citet{Burkert2020} relationship. We explore UDGs in stellar mass -- halo mass space further in Section \ref{sec:smhm}.

It is worth noting that further studies of UDGs in the NIHAO simulations have shown that NIHAO can model UDGs in group environments and at higher stellar masses \citep{Jiang2019, Cardona2020}. It is therefore likely that the full mass profile range that is reproducible by the NIHAO simulations is not fully captured by what we are plotting from \citet{DiCintio2017}. However, UDGs at higher stellar masses are expected to have higher mass profiles (see Table \ref{tab:comparison}). These higher stellar mass UDGs will therefore not affect the conclusions we draw from Figure \ref{fig:kong_compare} \textit{lower}. Additionally, some of the UDG sample used in \citet{DiCintio2017} includes galaxies that do not strictly meet the original \citet{vandokkum2015} UDG definition (i.e., they include galaxies with $R_{\rm e}<1.5$ kpc).

\subsection{UDGs on the Stellar Mass -- Halo Mass Relationship} \label{sec:smhm}

\begin{figure*}
    \centering
    \includegraphics[width=0.9 \textwidth]{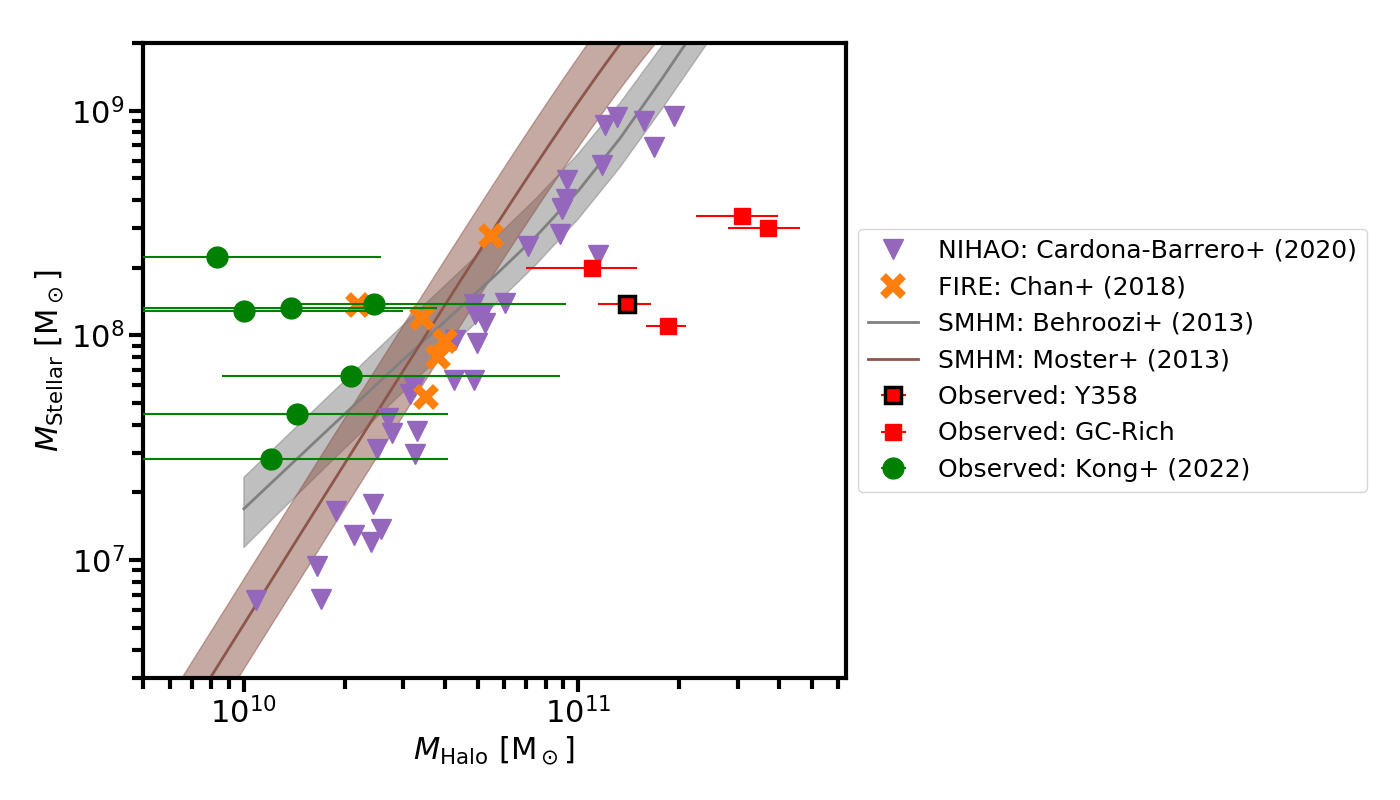}
    \caption{Stellar mass \textit{vs.} halo mass. We plot Y358 (black border) along with other GC-rich UDGs from \citet{Gannon2021b} as red squares. The isolated, gas-rich UDGs of \citet{Kong2022} UDGs are shown as green circles. The simulated FIRE UDGs are shown as orange crosses with NIHAO UDGs from \citet{Cardona2020}as purple triangles. Stellar mass -- halo mass relations are shown from \citet[grey line and shaded band]{Behroozi2013} and from \citet[brown line and shaded band]{Moster2013}. Both simulations create UDGs that generally follow conventional stellar mass -- halo mass relations. The observed GC-rich UDGs have halo masses more massive than either stellar mass -- halo mass relationship at fixed stellar masses. The observed isolated, gas-rich UDGs from \citet{Kong2022} have halo masses less massive than either stellar mass -- halo mass relationship at fixed stellar masses. Neither observed sample is reproduced by the FIRE or NIHAO simulations.}
    \label{fig:smhm}
\end{figure*}

\begin{table}
    \centering
    \begin{tabular}{llll}
    \hline
    Sample & n & $\Delta_{\rm B13}$ & $\Delta_{\rm M13}$ \\
    & & [dex] & [dex] \\ \hline
    Observed: Y358 & 1 &  0.49 & 0.54 \\
    Observed: GC-Rich & 5 & 0.53 & 0.63 \\
    Observed: Kong+ (2022) & 7 &  -0.37 & -0.38 \\
    NIHAO: Cardona-Barrero+ (2020) & 37 & 0.11 & 0.16 \\
    FIRE: Chan+ (2018) & 6 &  -0.03 & -0.01 \\ \hline
    \end{tabular}%
    \caption{A summary of the deviations from the stellar mass -- halo mass relationships of \citet{Behroozi2013} and \citet{Moster2013} for each sample using Equation \ref{eqtn:delta}. From left to right columns are: 1) the sample; 2) the number of objects in the sample; 3) $\Delta_{\rm B13}$, the average deviation calculated for \citet{Behroozi2013} and 4) $\Delta_{\rm M13}$, the average deviation calculated for \citet{Moster2013}. Note that Y358 is included in the calculation of the GC-rich statistics.}
    \label{tab:smhm}
\end{table}

In Figure \ref{fig:smhm} we further investigate observed UDGs \textit{vs.} the FIRE and NIHAO simulations by comparing them with the stellar mass -- halo mass relationship. Y358 and GC-rich UDGs from \citet{Gannon2021b} are plotted using halo mass measurements calculated from their GC-numbers \citep{Burkert2020}. Here we do not plot the UDGs PUDG\_S74 and PUDG\_R84, along with the GC-poor sample, as their exact GC counts with errors are unpublished, leading to an uncertain total halo mass. \citet{Kong2022} UDGs are plotted using the total halo mass coming from best fitting \citet{Read2016} halo profiles (see their table 1). The data for the \mbox{NIHAO} sample presented in \citet{Cardona2020} were attained from the corresponding author. The data for the FIRE UDGs are taken from their table 2 \citep{Chan2018}. Stellar mass -- halo mass relationships are plotted from the studies of \citet{Behroozi2013} and \citet{Moster2013}.

To quantify the deviation of each sample plotted in Figure \ref{fig:smhm} from established stellar mass -- halo mass relationships we define the quantity $\Delta_{\rm SMHM}$ for a sample of size $n$. This is the average logarithmic difference between the measured halo masses $M_{\rm halo, UDG}$ and the expected halo mass at the stellar mass of each UDG $ M_{\rm halo, SMHM}$ based on a stellar mass -- halo mass relationship.

\begin{equation}\label{eqtn:delta}
    \Delta_{\rm SMHM} = \frac{1}{n} \sum_{i=1}^{n} \log_{10}\ \frac{M_{\rm Halo, UDG}}{M_{\rm Halo, SMHM}}
\end{equation}

When using Equation \ref{eqtn:delta} for \citet{Behroozi2013} and \citet{Moster2013} we refer to it as $\Delta_{\rm B13}$ and $\Delta_{\rm M13}$ respectively. Positive values for $\Delta_{\rm SMHM}$ indicate that the sample resides in dark matter halos that are on average more massive than the stellar mass -- halo mass relationship. Negative values for $\Delta_{\rm SMHM}$ indicate that the sample resides in dark matter halos that are on average less massive than the stellar mass -- halo mass relationship. Values of $\Delta_{\rm SMHM}$ near zero indicate the sample obeys the relationship. We summarise the values of $\Delta_{\rm SMHM}$ in Table \ref{tab:smhm}. Note that 7 of the \citet{Cardona2020} UDGs are excluded from the calculation of $\Delta_{\rm B13}$ as they have stellar masses below the relationship's minimum value ($M_{\rm \star, Min} = 1.7 \times 10^{7}$~$\mathrm{M_{\odot}}$).

It is clear from Figure \ref{fig:smhm} and Table \ref{tab:smhm} that both simulated UDG samples largely follow known stellar mass -- halo mass relations (average $|\Delta_{\rm SMHM}|<0.2$~dex). This is less than the typical scatter (0.2 dex) in these relations. The only exception is the low mass (both stellar and total halo) end of the \citet{Cardona2020} data which does not follow \citet{Behroozi2013}. In contrast, both observational samples deviate strongly from both stellar mass -- halo mass relations. The isolated, gas-rich UDGs of \citet{Kong2022} reside in halos less massive than the stellar mass -- halo mass relation predicts for their stellar mass ($\Delta_{\rm B13} = -0.37$~dex; $\Delta_{\rm M13} = -0.38$~dex). This conclusion has been reached previously for a similar UDG sample by \citet{TrujilloGomez2022}. Note also that, it is unlikely that the low halo masses of these UDGs are caused by tidal stripping due to their isolated environments. The GC-rich UDG sample (which includes Y358) resides in halos more massive than the stellar mass -- halo mass relationship predicts for their stellar mass ($\Delta_{\rm B13} = 0.53$~dex; $\Delta_{\rm M13} = 0.63$~dex). Despite both FIRE and NIHAO reproducing the observed dynamical masses of GC-rich UDGs (Fig. \ref{fig:kong_compare}) they do not reproduce their inferred halo mass at their stellar mass.

\section{Extended Literature Sample} \label{sec:lit_sample}
For the remainder of this paper, we supplement our data for Y358 with data taken from the literature for spectroscopically studied UDGs. We take those galaxies from the literature that meet a UDG definition of $R_{\rm e} > 1.5$ kpc and $\langle\mu_{V}\rangle_{\rm e} > 24.7$ mag arcsec$^{-2}$. Our surface brightness criterion is simply that used in \citet{Gannon2021b}, $\langle\mu_{g}\rangle_{\rm e} > 25$ mag arcsec$^{-2}$, transformed into $V$-band with a colour of $V=g-0.3$. We have identified 21 galaxies in the literature meeting this definition with basic properties to place them in mass--radius--luminosity space. These properties (i.e., identifier, environment, distance, Mag., $\langle\mu_{V}\rangle_{\rm e}$, stellar mass, $R_{\rm e}$, recessional velocity, velocity dispersion and GC counts) are listed in Table \ref{tab:sample}. We include notes as to the construction of this sample in Appendix \ref{app:literature_sample}.


\subsection{UDG Spectroscopic Sample Biases} \label{sec:biases}
We note our UDG sample originates from a wide range of literature sources and therefore is not complete. We therefore briefly mention two obvious biases in the sample. Namely, UDGs in our literature sample tend to be 1) larger and 2) brighter in surface brightness than the broader UDG population.

\begin{figure}
    \centering
    \includegraphics[width = 0.98 \columnwidth]{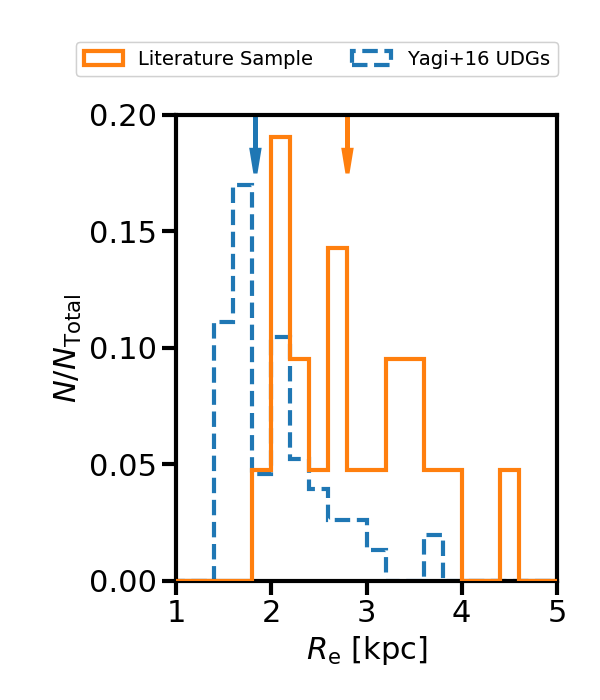}
    \caption{Normalised histograms of UDG circularised half-light radii. We plot our UDG sample ($N$ = 21; orange solid line) in comparison to the \citet{Yagi2016} $R$-band catalogue of Coma cluster objects that are UDGs ($N$ = 153; blue dashed line). Median values for each sample are indicated by arrows at the top of the plot. Our UDG sample has generally larger half-light radii than the Coma sample.}
    \label{fig:re_hist}
\end{figure}


In Figure \ref{fig:re_hist} we plot a histogram of UDG sizes for both our literature UDG sample and a subset of the $R$-band Coma cluster catalogue of \citet{Yagi2016} that are UDGs ($R_{\rm e}$ $> 1.5$ kpc and $\langle \mu_{R} \rangle_{\rm e} > 25$ mag arcsec$^{-2}$). We use this sample due to their likely association with the Coma Cluster which will decrease the uncertainty in their true size in comparison to a UDG sample of unknown distance. The use of a catalogue in a different filter band is expected to have only a small ($\approx10\%$) effect on half-light radii (see e.g., the UDG fitting in table 2 of \citealp{Saifollahi2022}) which is not large enough to affect our results. Performing a Kolmogorov–Smirnov test, it is highly unlikely that our UDG sample was randomly selected from the UDGs in the \citet{Yagi2016} catalogue ($p$ value = 0.005). Our literature sample is larger, with median half-light radius (2.8 kpc) larger than the \citet{Yagi2016} catalogue (median half-light radius 1.83 kpc).  


\begin{figure}
    \centering
    \includegraphics[width = 0.98 \columnwidth]{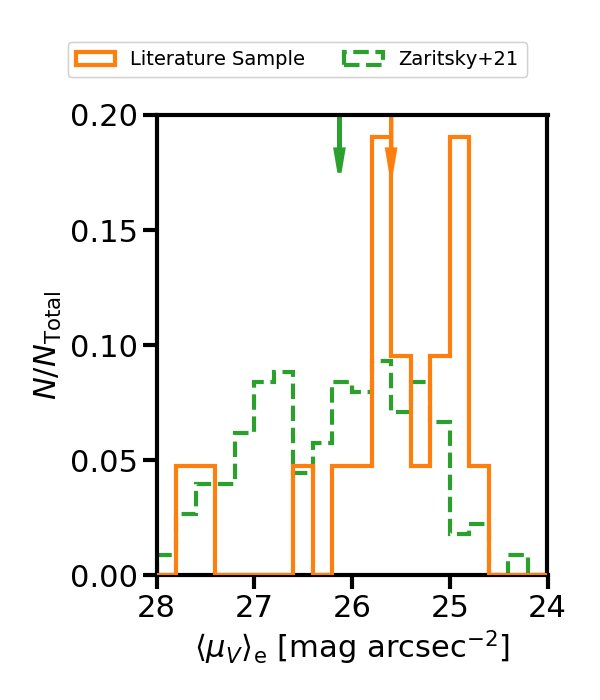}
    \caption{Normalised histograms of UDG surface brightnesses. We plot our UDG sample ($N$ = 21; orange solid line) in comparison to UDG candidates in the Stripe82 region ($N$ = 226; green dashed line; \citealp{Zaritsky2021}). Median values for each sample are indicated by arrows at the top of the plot. Our UDG sample has brighter surface brightnesses on average than UDGs in the Stripe82 region.}
    \label{fig:sb_hist}
\end{figure}


In Figure \ref{fig:sb_hist} we plot a histogram of the surface brightnesses of our literature UDG sample. We include for comparison UDG candidates from the Stripe 82 SMUDGes catalogue of \citet{Zaritsky2021}. Here we do not reuse the \citet{Yagi2016} catalogue due to the need for a common filter band to compare surface brightnesses. Additionally, the \citet{Zaritsky2021} catalogue provides the benefit of having UDGs across a full range of environments (field to cluster). For the \citet{Zaritsky2021} data we convert their measured central surface brightnesses to the average within the half-light radius using equation 11 of \citet{Graham2005} for comparison to our other data. We also correct this $g$-band catalogue into $V$-band using $V=g-0.3$. Performing a Kolmogorov–Smirnov test, it is unlikely that our UDG sample was randomly selected from the \citet{Zaritsky2021} catalogue ($p$ value = $0.012$). Our literature sample has a median surface brightness (25.6 mag arcsec$^{-2}$) brighter than the \citet{Zaritsky2021} catalogue (median surface brightness 26.1 mag arcsec$^{-2}$). We note that this is despite the \citet{Zaritsky2021} catalogue containing blue UDGs which will create a bias in their sample to be brighter due to their younger ages. 



\begin{figure}
    \centering
    \includegraphics[width = 0.98 \columnwidth]{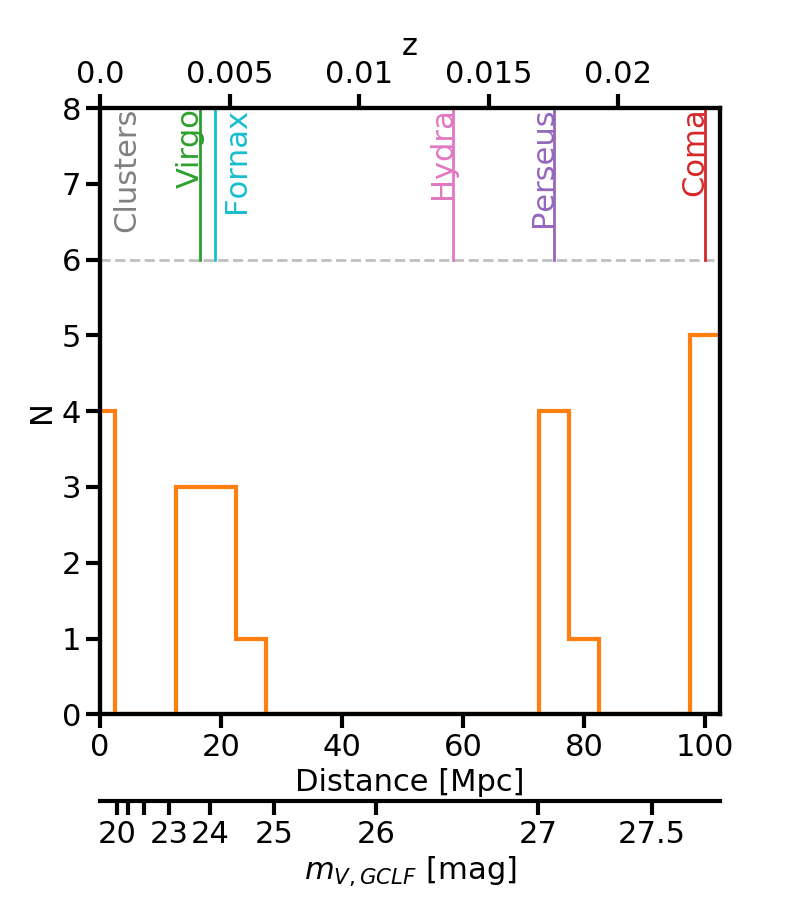}
    \caption{A histogram of UDG distances for UDGs from our literature sample. The x-axis shows the Distance (Mpc), redshift ($z$) and the apparent magnitude of the GC luminosity function peak. The positions of select clusters are given along the top of the plot (vertical coloured lines). To date, no UDGs have been targeted for deep spectroscopy at distances beyond 100 Mpc.}
    
    \label{fig:dist_hist}
\end{figure}

\begin{figure*}
    \centering
    \includegraphics[width = 0.98 \textwidth]{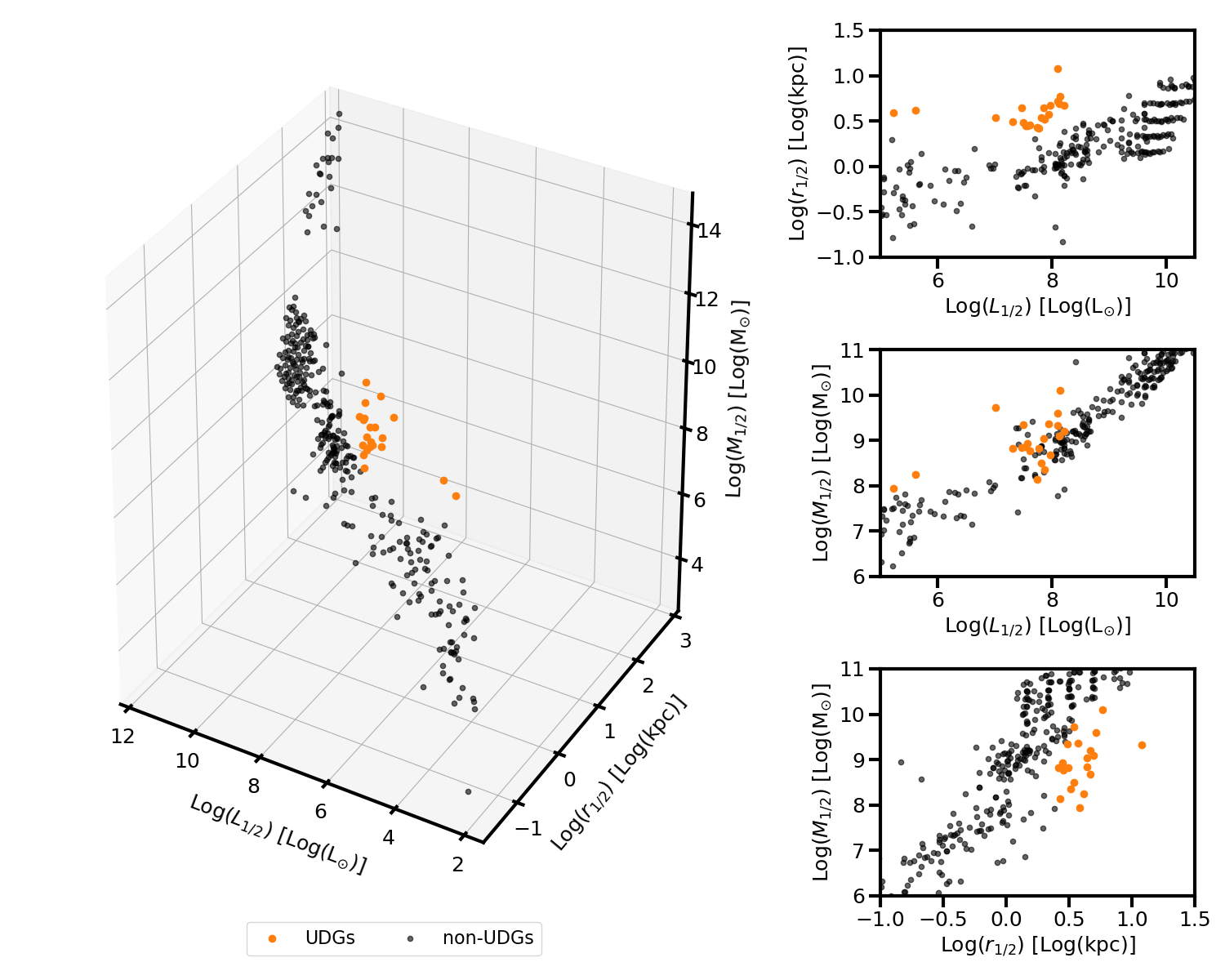}
    \caption{\textit{Left:} Mass--radius--luminosity space: half-light luminosity ($L_{1/2}$), half-light radius ($r_{1/2}$) and dynamical mass within the half-light radius ($M_{1/2}$). We project the plane and zoom around the location of UDGs on the \textit{Right}. From top to bottom these are the $L_{1/2}$ -- $r_{1/2}$, $L_{1/2}$ -- $M_{1/2}$ and $r_{1/2}$ -- $M_{1/2}$ projections of the plane. We establish the locus for non-UDGs using the data from \citet{Tollerud2011, Toloba2012, mcconnachie2012, Kourkchi2012} and \citet{forbes2018} (black points). See text for more details on these data. We also include our UDG sample on the plane (orange). UDGs are located off the locus for normal galaxies with the primary difference being their larger sizes.}
    \label{fig:fundamental_plane}
\end{figure*}

To further contextualise our literature sample we show a histogram of their distances in Figure \ref{fig:dist_hist}. We include the peak of the GC luminosity function at each distance based on an assumed peak of $M_{V} = -7.3$ \citep{Miller2007}. We also include a number of commonly studied clusters. To date, no UDGs have been targeted for deep spectroscopy at distances beyond 100 Mpc.


In order to best establish dark matter halo profile parameters, accurate radial mass profiles are required. For UDGs, a thorough exploration of their likely cored dark matter halos will require observations to be made beyond the dark matter core radius ($\sim 5-10$ kpc). The current single mass measurements available for many UDGs are insufficient to truly establish dark matter halo parameters due to degeneracies in their comparison to theoretical halo mass profiles \citep{Gannon2021}. GCs pose a promising avenue to get larger radius mass estimates for UDGs to help probe their halo profile (e.g., \citealp{Gannon2020}). Importantly this suggests UDG observational efforts should be focused on those candidates nearby enough to allow spectroscopic studies of their GC system.


\section{Discussion: Mass--Radius--Luminosity Space} \label{sec:MRL}

In Figure \ref{fig:fundamental_plane} we place UDGs in mass--radius--luminosity space, an altered form of the fundamental plane for pressure-supported systems. We establish the locus traced by non-UDGs using data from \citet{Tollerud2011, Toloba2012, mcconnachie2012, Kourkchi2012} and \citet{forbes2018}. For \citet{Toloba2012} galaxies we convert half-light radii into physical units using an assumed Virgo cluster distance of 16.5 Mpc. We place \citet{Kourkchi2012} galaxies on the plane using a correction of $V = F814W + 1$. We place \citet{mcconnachie2012}\footnote{January 2021 public version} galaxies on the plane using their given Vega magnitudes. We then place our literature UDG sample on the plane to examine their location. We convert magnitudes into solar units assuming $M_{V, \mathrm{Sun}}$ = 4.8 \citep{Willmer2018} and dynamical masses are calculated using Equation \ref{eqtn:wolf}. Galaxies that fit the UDG definition in the non-UDG samples (e.g., the Sagittarius dSph appears in both \citealp{mcconnachie2012} and \citealp{forbes2018}) are removed before plotting as they are included in our literature sample.

We note two UDGs plotted, Andromeda XIX and Antlia II, have half-light luminosities noticeably less bright than the remaining population. These galaxies have measured velocity dispersions only due to their extremely close distances (i.e., both are in the Local Group) which allows their stars to be resolved. We note that there exists a continuum of galaxies of large size and varying luminosity between these galaxies and the remaining UDGs on the relation (see e.g., table 2 of \citealp{Karachentsev2017}). The empty region between these galaxies and the remainder of our sample is simply a side effect of our bias to higher surface brightness objects. For the remainder of our discussion we will focus on the higher-luminosity objects more readily studied. 

Interestingly, UDGs reside in a region of parameter space largely separate from the locus of non-UDGs. Their main difference is simply their larger half-light radii, with dynamical masses and luminosities similar to the locus of non-UDGs (the empty region between these two populations in half-light radii on Figure \ref{fig:fundamental_plane} exists due to selection effects). An unfortunate corollary of UDGs inhabiting an entirely new parameter space is that, at fixed luminosity, UDG masses cannot be estimated from luminosity and radius information alone. This will hamper efforts to perform statistical estimations of UDG masses based on their photometric properties (e.g., \citealp{Zaritsky2017, Lee2020}). 

The similarity in luminosities between UDGs and non-UDGs have led many to suggest they may simply be an extension of the dwarf galaxy population to larger sizes (see e.g., \citealp{Conselice2018}). These UDGs could be `puffy' dwarf galaxies formed through conventional pathways (e.g. \citealp{Amorisco2016, DiCintio2017, Rong2017, Tremmel2020}). A likely example of these are the GC-poor, cluster UDGs plotted in Figure \ref{fig:kong_compare} which are plausibly the result of the transformation of extended, star forming field dwarfs (e.g., \citealp{Grishin2021}). In mass--radius--luminosity space `puffy dwarf' UDGs are expected to have similar luminosities, larger radii and dynamical masses only slightly larger than their non-UDGs of similar luminosity. This is reflective of their similar dark matter halos. Much of our literature sample has mass, radius and luminosity compatible with this expectation for a `puffy dwarf' formation scenario. 


We caution that this expectation for `puffy dwarf' UDGs may be over-simplified. \citet{Kadowaki2021} found that the dynamical masses of UDGs measured with increasingly large radii likely correspond to increasingly massive dark matter halos (see their appendix A). In this framework, many of the UDGs in our sample may have dynamical masses corresponding to dark matter halos more massive than non-UDGs at similar luminosity. If this is the case, these UDGs cannot be explained by `puffy dwarf' formation scenarios due to their massive dark matter halos. We do note however, that due to the bias to UDGs with larger half-light radius in our sample, we expect a greater fraction of our sample to be massive halo UDGs than the UDG population as a whole.

Finally, we suggest Figure \ref{fig:fundamental_plane} is a fundamental empirical plot that should be reproduced by galaxy formation simulations focusing on UDGs. Particularly, many simulations currently have difficulty reproducing the full range of dwarf galaxy sizes. For example the ROMULUS-C simulations currently form the majority of their dwarfs in the UDG stellar mass regime as UDGs (see \citealp{Tremmel2020} table 1). In addition, UDG studies using the Illustris simulations have to assign their UDG candidates sizes due to limitations of their simulations \citep{Carleton2019, Sales2020}. Simulations reproducing the full range of galaxy sizes and masses in the UDG luminosity regime will be crucial to developing a theoretical understanding of their formation.

\section{Conclusions} \label{sec:conclusions}
In this work we have added Keck/KCWI spectroscopy of a GC-rich, Coma cluster UDG, Y358, to the literature.  We then create a literature sample of UDGs that have been studied spectroscopically, placing them in mass--radius--luminosity space for pressure-supported galaxies. Our main conclusions are as follows:

\begin{itemize}
    \item We measure a recessional velocity (7969 $\pm$ 2 km s$^{-1}$) and velocity dispersion (19 $\pm$ 3 km s$^{-1}$) for Y358. The recessional velocity confirms its association with the Coma cluster. This association formalises the distance of Y358 and its status as a UDG. 
    
    \item We calculate a dynamical mass for Y358 and compare it to cuspy and cored dark matter halo profiles with total mass inferred from its GC count. Under the assumption that the total halo mass from GC counts is correct, Y358 likely resides in a cored dark matter halo. 

We supplement our dynamical mass measurement for Y358 with others from \citet{Gannon2021b}. We then compare to the best-fitting dark matter mass profiles for isolated, gas-rich UDGs from \citet{Kong2022}. We also compare the \citet{Kong2022} measurements to simulations of UDG formation from the NIHAO suite \citep{DiCintio2017} and the FIRE suite \citep{Chan2018}. We find:
    
    \item The GC-poor UDGs may reside in a dark matter halo of similar radial profile to the isolated, gas-rich UDGs of \citet{Kong2022}, suggesting an evolutionary connection may exist between the two populations. Dynamical mass measurements made for GC-rich UDGs are sufficiently high to exclude them residing in dark matter halos similar to the isolated, gas-rich UDGs. We suggest it is unlikely that GC-rich UDGs represent an evolved population of isolated, gas-rich UDGs.
    
    \item Both the simulations of \citet{DiCintio2017} and \citet{Chan2018} produce mass profiles for UDGs that are too massive when compared to the isolated, gas-rich UDGs of \citet{Kong2022}. This is unexpected as the \citet{Kong2022} samples have the properties (i.e., stellar mass, total halo mass, environment and gas-richness) most resembling the UDG in their simulations. The simulated mass profiles are instead more consistent with GC-rich UDGs.
    
\item We find that although FIRE and NIHAO simulations cover the stellar and halo mass range of GC-rich UDGs, they cannot reproduce their observationally-estimated halo masses at the same stellar mass. This is perhaps not unexpected given that the simulations are for isolated UDGs, whereas our observed UDGs are located in groups and clusters where additional environmental effects may play a role in their evolution.

\item 
    
We then gather a literature sample for all galaxies meeting our UDG definition with spectroscopic velocity dispersions. We find two biases in this sample: 
\begin{enumerate}
        \item The UDGs in our literature sample are on average larger than the population as a whole.
        \item The UDGs in our literature sample have brighter surface brightness than the population as a whole.
\end{enumerate}       
Both of these need to be kept in mind when considering UDG formation scenarios from current observational data. We then place our UDG sample in mass--radius--luminosity space, examining their location. We find:

    \item UDGs are located at a similar locus to non-UDGs of similar luminosity with the primary difference being their increased half-light radius. This supports notions that some UDGs are simply `puffy dwarfs' with extended sizes driven by known physical processes.
    
    \item UDGs' dynamical masses within their large radii may indicate massive dark matter halos not expected in a `puffy dwarf' formation scenario. As our UDG sample is biased to the largest systems, we suggest a greater fraction of UDGs in our sample may be massive halo UDGs than the population as a whole.

\end{itemize}

\section*{Acknowledgements}

We thank the anonymous referee for their considered reading of this work and comments which have improved its quality. We thank S. Cardona-Barrero for providing us with their NIHAO data. We thank A. Alabi for aiding with the KCWI observations. JSG thanks R. Abraham, M. L. Buzzo and S. Janssens for insightful conversations throughout the creation of this work. Some of the data presented herein were obtained at the W. M. Keck Observatory, which is operated as a scientific partnership among the California Institute of Technology, the University of California and the National Aeronautics and Space Administration. The Observatory was made possible by the generous financial support of the W. M. Keck Foundation. The authors wish to recognise and acknowledge the very significant cultural role and reverence that the summit of Maunakea has always had within the indigenous Hawaiian community.  We are most fortunate to have the opportunity to conduct observations from this mountain. AJR was supported as a Research Corporation for Science Advancement Cottrell Scholar. JSG acknowledges financial support received through a Swinburne University Postgraduate Research Award throughout the creation of this work. AFM acknowledges support from the Severo Ochoa Excellence scheme (CEX2019-000920-S) and from grant PID2019-107427GB-C32 from the MCIU of Spain.

\section*{Data Availability}
The KCWI data presented are available via the Keck Observatory Archive (KOA): \url{https://www2.keck.hawaii.edu/koa/public/koa.php} 18 months after observations are taken. The literature sample discussed is included in Appendix \ref{app:literature_sample}.


\bibliographystyle{mnras}
\bibliography{bibliography} 



\appendix

\section{Literature UDG Data} \label{app:literature_sample}

In this Appendix we present the literature sample of spectroscopically analysed UDGs used in Section \ref{sec:lit_sample}. Lettering in the notes below corresponds to the superscripts in Table \ref{tab:sample}.

\subsubsection{Y358}
\textit{Notes:} $a$ = Calculated from the absolute magnitude assuming $M_{\star} / L_{V} =2$ and $M_{\odot, V}$ = 4.8 \citep{Wilmer2018}.  $b$ = Circularised using literature $b/a$ (0.83; \citealp{vanDokkum2017}). Data sources: This work, \citet{vanDokkum2017} and \citet{Lim2018}.  

\subsubsection{VCC~1287}
\textit{Notes:} $a$ = This galaxy is identified as NGVSUDG-14 in \citet{Lim2020}. $b$ = It is unclear which filter band this is in, although $g$-band seems likely from the context. We therefore transform it into $V$-band using $V$ = $g - 0.3$. Data sources: \citet{Beasley2016, Gannon2020, Gannon2021, Lim2020}. 

\subsubsection{DGSAT~I}
\textit{Notes:} $a$ = It is located in the Pisces--Perseus supercluster and could potentially be a `backsplash' galaxy \citep{MartinezDelgado2016, Papastergis2017}. $b$ = Calculated using the properties listed in Table 2 of \citet{MartinezDelgado2016} and equation 11 of \citet{Graham2005}. $c$ = Circularised using literature $b/a$. $d$ = Note that some of these GC's may be overluminous \citep{Janssens2022}. Data sources: \citet{MartinezDelgado2016, MartinNavarro2019, Janssens2022}. 

\subsubsection{Dragonfly~44}
\textit{Notes:} $a$ = Although in the direction of the Coma cluster ``it is unclear whether Dragonfly~44 is in a cold clump that is falling into the cluster, a filament, or a structure that is unrelated to Coma'' \citep{vanDokkum2019b}. $b$ = Calculated using the properties listed in Table 1 of \citet{vanDokkum2017} at a distance of 100 Mpc and equation 11 of \citet{Graham2005}. $c$ = Converted using $V_{\rm R} = c \times \ln{1+z}$ from the redshift listed in footnote 6 of \citet{vanDokkum2017}. $d$ = Note the $N_{\rm GC}$ quoted in the abstract is slightly different to this value. Here we use the value from Table 1 of \citet{vanDokkum2017}. $e$ = Although see \citet{Saifollahi2021} for a differing view of the GC richness of Dragonfly~44. Data sources: \citet{vanDokkum2016, vanDokkum2017, vanDokkum2019b, Gannon2021}.

\subsubsection{DFX1}
\textit{Notes:} $a$ = Calculated using the properties listed in Table 1 of \citet{vanDokkum2017} at a distance of 100 Mpc and equation 11 of \citet{Graham2005}. $b$ = Converted using $V_{\rm R} = c \times \ln{1+z}$ from the redshift listed in section 2.1 of \citet{vanDokkum2017}. $c$ = it is unclear if this is also effected by the barycentric correction issue described in footnote 16 of \citet{vanDokkum2019b}. Data sources: \citet{vanDokkum2017, Gannon2021}.

\subsubsection{NGC~5846\_UDG1}
\textit{Notes:} $a$ = This galaxy is referred to as MATLAS-2019 in the MATLAS dwarf galaxy catalog \citep{Habas2020, Muller2020, Muller2021}. $b$ = Calculated using the properties listed in Table 1 of \citet{Forbes2019} and equation 11 of \citet{Graham2005}. Transformed from $g$-band using $V = g-0.3$. $c$ = We preference these values over those reported in \citet{Muller2020} due to the greater instrumental resolution of Keck/KCWI over VLT/MUSE. $d$ = We preference these values over those reported in \citet{Muller2021} due to the deeper \textit{HST} data used. Data sources: \citet{Muller2020, Muller2021, Forbes2021, Danieli2022}.

\subsubsection{VLSB-B}

\textit{Notes:} $a$ = This galaxy is identified as NGVSUDG-11 in \citet{Lim2020}. $b$ = Vega magnitude. $c$ = Circularised using literature ellipticity. $d$ = This is a GC system velocity dispersion that we assume is equivalent to the stellar velocity dispersion of the galaxy based on the evidence for this assumption in \citet{Forbes2021}. Data sources: \citet{Toloba2018, Lim2020}. 

\subsubsection{VLSB-D}

\textit{Notes:} VLSB-D has an elongated structure and velocity gradient \citep{Toloba2018} that suggests it is undergoing tidal stripping. $a$ = This galaxy is identified as NGVSUDG-04 in \citet{Lim2020}. $b$ = Vega magnitude. $c$ = Circularised using literature ellipticity. $d$ = This is a GC system velocity dispersion that we assume is equivalent to the stellar velocity dispersion of the galaxy based on the evidence for this assumption in \citet{Forbes2021}. Data sources: \citet{Toloba2018, Lim2020}

\subsubsection{VCC~615}
\textit{Notes:} \textit{Notes:} $a$ = This galaxy is identified as NGVSUDG-A04 in \citet{Lim2020}. $b$ = Vega magnitude. $c$ = Circularised using literature ellipticity. $d$ = This is a GC system velocity dispersion that we assume is equivalent to the stellar velocity dispersion of the galaxy based on the evidence for this assumption in \citet{Forbes2021}. Data sources: \citet{Toloba2018, Lim2020, Mihos2022}. 

\subsubsection{UDG1137+16}
\textit{Notes:} UDG1137+16 has a disturbed morphology making it likely it is undergoing stripping \citep{Gannon2021}. $a$ = See also \citet{Muller2018} who refer to this galaxy as `dw1137+16'. Transformed to $V$-band using stated $g-r$ colour and $V=g-0.3$. $c$ = Calculated using the properties listed for the single S\'ersic fit in Table 1 of \citet{Gannon2021} and equation 11 of \citet{Graham2005}. Data source: \citet{Gannon2021}.

\subsubsection{PUDG-R15}
\textit{Notes:} $a$ = Transformed from $g$-band using $V = g-0.3$. $b$ = Circularised using literature $b/a$. Data source: \citet{Gannon2021b}.

\subsubsection{PUDG-R16}
\textit{Notes:} $a$ = Transformed from $g$-band using $V = g-0.3$. $b$ = Circularised using literature $b/a$. Data source: \citet{Gannon2021b}.

\subsubsection{PUDG-S74}
\textit{Notes:} $a$ = Transformed from $g$-band using $V = g-0.3$. $b$ = Circularised using literature $b/a$. Data source: \citet{Gannon2021b}. 

\subsubsection{PUDG-R84}
\textit{Notes:} $a$ = Transformed from $g$-band using $V = g-0.3$. $b$ = Circularised using literature $b/a$. Data source: \citet{Gannon2021b}.

\subsubsection{NGC~1052-DF2}
\textit{Notes:} This galaxy has both an anomalous star cluster system \citep{vanDokkum2018b, Shen2021} and an abnormally low velocity dispersion \citep{vanDokkum2018, Danieli2019}. There is also evidence it may be undergoing a tidal interaction (\citealp{ Keim2021}, although see \citealp{Montes2021}). We do however note there is some evidence for rotation in NGC~1052-DF2 which may help alleviate the paucity of dark matter implied by its low velocity dispersion \citep{Lewis2020, Montes2021}. $a$ = This is also catalogued as [KKS2000]04 in \citet{Karachentsev2000}. $b$ = While there existed some initial controversy over the distance of NGC~1052-DF2 (see e.g., \citealp{Trujillo2019, Monelli2019}) we believe the deeper data reported in \citet{Shen2021} resolved this issue. We note however, despite an established distance this does not fully establish an environmental association for NGC~1052-DF2 (see e.g., Fig. 5 of \citealp{Shen2021}). The possibility exists that NGC~1052-DF2 was part of the NGC~1052 group but now resides outside of the group as a consequence of its formation (e.g., \citealp{vanDokkum2022}). $c$ = Calculated using the properties listed for the single S\'ersic fit in Table 2 of \citet{Cohen2018} and equation 11 of \citet{Graham2005}. $d$ = We preference these values over those reported in \citet{Emsellem2019} due to the greater instrumental resolution of Keck/KCWI over VLT/MUSE. $e$ = Here we use the value of GCs in the roughly expected GC luminosity function window as reported by \citet{Shen2021}. This value excludes the brighter sub-population. Data sources: \citet{Danieli2019, Shen2021}.

\subsubsection{Sagittarius dSph}
\textit{Notes:} Note this galaxy is known to be tidally disrupting around the Milky Way \citep{Ibata2001} $^a$ Calculated using the properties listed in Table 1 of \citet{forbes2018} and equation 12 of \citet{Graham2005}. Data sources: \citet{mcconnachie2012, Karachentsev2017, forbes2018}.

\subsubsection{Andromeda XIX}
\textit{Notes:} Note this galaxy is likely affected by tidal processes interacting with the nearby M31 \citep{Collins2020}. $a$ = Calculated using the properties listed in Table 3 of \citet{Collins2020} and equation 12 of \citet{Graham2005}. Due to the extremely diffuse nature of this object this value is highly uncertain. Data sources: \citet{Martin2016, Collins2020, Gannon2021}.

\subsubsection{Antlia II}
\textit{Notes:} Dynamical modelling suggests that a combination of a cored dark matter profile and tidal stripping may explain the properties of this UDG \citep{torrealba2019}. $a$ = Vega magnitude. $b$ Due to the extremely faint nature of Antlia II this value is highly uncertain. Data sources: \citet{mcconnachie2012, torrealba2019}

\subsubsection{WLM}
\textit{Notes:} WLM is gas-rich and undergoing active star formation \citep{Leaman2009}. $a$ = calculated from given m - M. $b$ = Vega magnitude. $^c$ Calculated using equation 12 of \citet{Graham2005}. $d$ = Calculated from $V$-band magnitude assuming $M_{\star}/L_{V}$ = 2. Data sources: \citet{mcconnachie2012, forbes2018}

\subsubsection{J125929.89+274303.0}
\textit{Notes:} $a$ = Converted from $R$-band using $V = R +0.5$ (based on Virgo dE's; \citealp{vanZee2004}) Data sources: \citet{Chilingarian2019, Gannon2021}.

\subsubsection{J130026.26+272735.2}
\textit{Notes:} $a$ = Converted from $R$-band using $V = R +0.5$ (based on Virgo dE's; \citealp{vanZee2004}) Data sources: \citet{Chilingarian2019, Gannon2021}.

\begin{landscape}

\begin{table}
\centering
\begin{tabular}{llllllllll}
\hline
Name & Env. & $D$ & $M_{V}$ & $\langle\mu_{V}\rangle_{\rm e}$ & $M_\star$ & $R_{\rm e}$ & $V_{\rm R}$ & $\sigma$ & $N_{\rm GC}$ \\
 &  & {[}Mpc{]} & {[}mag{]} & {[}$\mathrm{mag\ arcsec^{-2}}${]} & {[}$\times\ 10^{8}\ \mathrm{M_{\odot}}${]} & {[}kpc{]} & {[}km $\mathrm{s^{-1}}${]} & {[}km $\mathrm{s^{-1}}${]} &  \\ \hline
Y358 & Cluster \{Coma\} & 100 & $-$14.8 & 25.6 & 1.38$^a$ & 2.1$^b$ & 7969 (2) & 19 (3) & 28 (5.3) \\
VCC 1287$^a$ & Cluster \{Virgo\} & 16.5 & $-$15.6 & 25.71$^b$ & 2 & 3.3 & 1116 (2) & 19 (6) & 22 (8) \\
DGSAT I & Field?$^a$ & 78 & $-$16.3 & 25.6$^b$ & 4 & 4.4$^c$ & 5439 (8) & 56 (10) & 12 (2)$^d$ \\
Dragonfly 44 & Cluster?$^a$ \{Coma\} & 100 & $-$16.2 & 25.7$^b$ & 3 & 3.9 & 6324$^c$ & 33 (3) & 76 (18)$^{d,e}$ \\
DFX1 & Cluster \{Coma\} & 100 & $-$15.8 & 25.5$^a$ & 3.4 & 2.8 & 8107$^b$ & 30 (7)$^c$ & 63 (17) \\
NGC 5846\_UDG1$^a$ & Group \{NGC 5846\} & 26.5 & $-$15.0 & 25.2$^b$ & 1.1 & 2.14 & 2167 (2)$^c$ & 17 (2)$^c$ & 54 (9)$^d$ \\
VLSB-B$^a$ & Cluster \{Virgo\} & 16.5 & $-$13.5$^b$ & 27.5 & 0.06 & 2.6$^c$ & 24.9 $(^{+22.3}_{-36.2})$ & 47 (+53, -29)$^d$ & 26.1 (9.9) \\
VLSB-D$^a$ & Cluster \{Virgo\} & 16.5 & $-$16.2$^b$ & 27.6 & 0.79 & 9.0$^c$ & 1033.8 $(^{+5.9}_{-5.5})$ & 16 (+6, -4)$^d$ & 13 (6.9) \\
VCC 615$^a$ & Cluster \{Virgo\} & 17.7 & $-$14.7$^b$ & 25.8 & 0.21 & 2.3$^c$ & 2094.0 $(^{+14.9}_{-13.0})$ & 32 (+17, -10)$^d$ & 30.3 (9.6) \\
UDG1137+16$^a$ & Group \{UGC 6594\} & 21.1 & $-$14.65$^b$ & 26.55$^{b,c}$ & 1.4 & 3.3 & 1014 (3) & 15 (4) & - \\
PUDG-R15 & Cluster \{Perseus\} & 75 & $-$15.65$^a$ & 24.83$^a$ & 2.59 & 2.46$^b$ & 4762 (2) & 10 (4) & - \\
PUDG-R16 & Cluster \{Perseus\} & 75 & $-$15.9$^a$ & 25.4$^a$ & 5.75 & 3.51$^b$ & 4679 (2) & 12 (3) & - \\
PUDG-S74 & Cluster \{Perseus\} & 75 & $-$16.49$^a$ & 24.82$^a$ & 7.85 & 3.52$^b$ & 6215 (2) & 22 (2) & - \\
PUDG-R84 & Cluster \{Perseus\} & 75 & $-$15.4$^a$ & 24.68$^a$ & 2.20 & 1.97$^b$ & 4039 (2) & 19 (3) & - \\
NGC1052-DF2$^a$ & Group \{NGC 1052\}$^b$ & 22.1 (1.2)$^b$ & $-$15.3 & 24.8$^c$ & 2 & 2 & 1805 (1.1)$^d$ & 8.5 (+2.3, -3.1)$^d$ & 7.1 (+7.33, -4.34)$^e$ \\
Sagittarius dSph & Group \{Local\} & 0.02 & $-$15.5 & 25.13$^a$ & 1.32 & 2.6 & 140 (2) & 11.4 (0.7) & 8 \\
Andromeda XIX & Group \{Local\} & 0.93 & $-$10 & $\sim$ 31$^a$ & 0.0079 & 3.1 & -109 (1.6) & 7.8 (+1.7, -1.5) & - \\
Antlia II & Group \{Local\} & 0.132 & $-$9.03$^a$ & $\sim$ 31.9$^b$ & 0.0088 & 2.9 & 290.7 (1.5) & 5.71 (1.08) & - \\
WLM & Group \{Local\} & 0.93$^a$ & $-$14.25$^b$ & 26.16$^c$ & 0.41$^d$ & 2.34 & -130 (1) & 17.5 (2) & 1 \\
J125929.89+274303.0 & Cluster \{Coma\} & 100 & $-$14.88 $^a$ & 25.17 $^a$ & 1.12 & 2.1 & 4928 (4) & 21 (7) & - \\
J130026.26+272735.2 & Cluster \{Coma\} & 100 & $-$16.27 $^a$ & 24.83 $^a$ & 1.56 & 3.7 & 6939 (2) & 19 (5) & - \\
\hline
%
%
\end{tabular}
\caption{Rows from left to right are: 1) Name, 2) Environment \{Name\}, 3) Distance - although note this is frequently the assumed distance, 4) $V$-band absolute magnitude, 5) Average $V$-band surface brightness within the half-light radius, 6) Stellar mass, 7) 2D circularised half-light radius, 8) Recessional velocity, 9) Velocity dispersion from stars or GCs and 10) GC system number. When relevant errors are given in (brackets) after values. Values unknown are indicated with a ``-". Notes on data are included with superscript letters. }
\label{tab:sample}
\end{table}
\end{landscape}

\bsp	
\label{lastpage}
\end{document}